# Dynamically configurable successively switchable multispectral plasmon induced transparency

Jietao Liu,[1,2]* Ioannis Papakonstantinou,[2] Haifeng Hu,[3] and Xiaopeng Shao[1]

[1]*School of Physics and Optoelectronic Engineering, Xidian University, Xi'an, 710071, China*
[2]*Photonic Innovations Lab, Department of Electronic and Electrical Engineering, University College London, London, WC1E 7JE, UK*
[3]*College of Information Science and Engineering, Northeastern University, Shenyang, 110819, China*
*Corresponding author: liujt@xidian.edu.cn*

**Plasmon-induced-transparency (PIT) in nanostructures has been intensively investigated, however, no existing metasurface nanostructure exhibits all-optically tunable properties, where the number of transparency windows can be tuned successively, and switched to 'off-state'. Here, we theoretically investigate and demonstrate dynamically tunable, multichannel PIT at optical frequencies. The in-plane destructive interference between bright and dark dipolar resonances in coupled plasmonic nanobar topologies is exploited to produce tunable PIT with unique characteristics. In particular, we demonstrate sequential polarization-selective multispectral operation whereby the number of PIT channels can be varied successively from '3' to '0'. The results provide a promising route for active manipulation of PIT and show potential applications for multifunctional dynamic nanophotonics devices.**

The analogue of Electromagnetic Induced Transparency (EIT) in plasmonic systems known as PIT, has led to intriguing new optical effects and has sparked increased interest in applications such as slow light, sensing, optical filters, enhanced nonlinear effects, optical information processing and optical switches [1-6]. The principle of PIT relies on the destructive interference between two resonant transition pathways, a direct excitation of a dipole-allowed bright-mode, and an indirect excitation of a metastable dark-mode, rendering a transparency window at selective resonant frequencies, [1-5]. A number of advantages are associated with PIT compared to its quantum counterpart EIT, including room-temperature operation, wide bandwidth, compactness and ease of integration with nanophotonic circuits, [6-10].

While some work exhibiting multispectral PIT by precise control of symmetry breaking or complex arrangement of multiple complex nanostructures has been reported in the literature, [11-15], a system with all-optically tunable, multispectral PIT, where the number of transparency windows can be switched continuously, is still missing. Meanwhile, there is urgent demand for compact, efficient, dynamically controlled PIT in numerous applications such as optical filtering, sensing, multichannel information processing and others [16-20]. Here we propose the first, to the best of our knowledge, 'all-in-one' multispectral PIT system by exploiting the coupled resonances in carefully designed plasmonic nanobar array topologies. Remarkably, we observe distinct features in PIT lineshape that give rise to dynamic all-optical switching of PIT channels (numbers of transparency-windows) from no PIT to one-channel PIT, to two-channel PIT, and even triple-PIT by simply adjusting the polarization angle of the incident optical field.

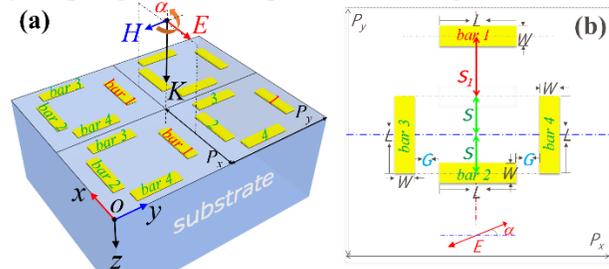

**Fig. 1.** (a) Isometric and (b) top view of schematic of the simulated PIT system consisting of a periodic gold nanoantenna topology, arranged on a square lattice.

The schematic of the plasmonic nanobar topology investigated in this study is illustrated in Figure 1. Each Unit cell comprises a pair of parallel metal strips (nanobars 1 and 2) working as an optical dipole nanoantenna for optical field polarization along *x*-direction, and a second pair of parallel metal strips (nanobars 3 and 4) vertically oriented along the *y*-direction. The geometrical parameters (width *W*, length *L* and thickness $t_m$) of the nanobars are taken to be *W* = 40 nm, *L* = 200 nm and $t_m$ = 20 nm, respectively. All nanobars have identical geometry and are made of gold. The nanobars are separated by variable gaps *S*, $S_1$ and *G* defined in Fig. 1(b), while unit cells are arranged on a square lattice with pitch $P_x$ = $P_y$ = 600 nm in

$x$ and $y$ correspondingly. Finally, the plasmonic nanobar system is assumed to rest on a silica substrate. Linearly polarized light is incident at a normal angle and with an angle of polarization $\alpha$. Full-field electromagnetic calculations were performed using Finite Difference Time Domain method. The permittivity, $\varepsilon_m$, of Au was taken from Johnson and Christie [21], while the refractive index of $SiO_2$ was fixed at 1.45.

A linearly-polarized optical field is incident with its electric component initially along the $x$-axis. Because of their orientation, the two nanobars 1 and 2 are strongly coupled to the incident field and act as dipole antennas supporting spectrally broad resonance. For specific wavelengths, these dipole antennas radiate to free-space with a large scattering cross-section giving rise to resonances in the transmission spectra (bright modes); the second nanobar pair, 3 and 4, is responsible for the excitation of non-radiative dark modes through the near-field interaction and coupling with the fields of the bright mode, since they cannot be excited directly by the external field.

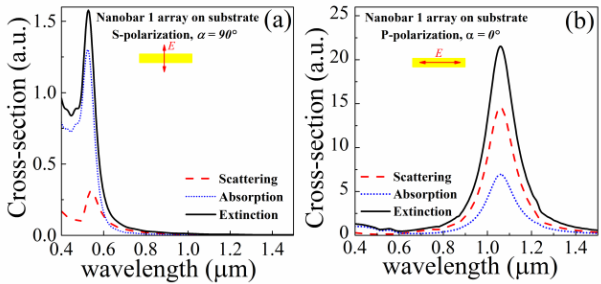

**Fig. 2.** Normalized extinction, scattering, and absorption cross-sections (black, red-dashed, and blue-dotted lines, respectively) for an individual array of nanobars 1 (or 2) on substrate (a) for vertical polarization $\alpha=90°$, and (b) for horizontal polarization $\alpha=0°$.

To elucidate these points more clearly, the resonance and scattering properties for isolated nano-antenna arrays are investigated firstly. The normalized extinction, scattering, and absorption cross-sections (black, red-dashed, and blue-dotted lines, respectively) when singling out the response from the periodic array of bars 1 are shown in Fig. 2(a) for vertical polarization ($\alpha=90°$), and Fig. 2(b) for horizontal polarization ($\alpha=0°$). The radiation properties of the nanobars are polarization dependent, and the results in figure 2 show stronger (one order of magnitude) scattering radiative feature at resonance for the nanobars that are oriented parallel to the incident polarization (1 and 2) than the orthogonal nanobars (3 and 4). The Lorentzian like linewidth in figure 2(b) for the nanobar antenna is characteristic of a radiative dipole resonance.

Figure 3(a) shows the transmission response of the individual nanobars as well as of the parallel (1 and 2) and orthogonal (3 and 4) nanobar pairs. The results for the individual nanobars (dashed lines) agree with the scattering cross-section in figure 2 as they should [22]. The results for the pairs (solid lines) on the other hand, indicate that only weak coupling between the nanobars within each pair is present. This is because of the relatively large separation between them, which results in their transmission to only slightly be increased and spectrally be shifted compared with the individual units. The situation changes dramatically when more complex configurations come into play, as shown in figure 3(b). The most notable effect is the emergence of a two-channel PIT (solid black line) when all four nanobars are present for appropriate values of the gap $G$ and the shifts $S$ and $S_1$ ($G$ = 12 nm, $S$ = 100 nm and $S_1$ = 170 nm).

By selectively removing first nanobar 1 and then nanobar 2, the individual contributions from the dark (blue dashed line) and the bright (red dashed line) modes can be extracted. Nanobar 1 is sufficiently decoupled from 3 and 4 and virtually only the bright mode is excited. When nanobar 2 is introduced, the excitation of the otherwise forbidden dark mode is allowed and a broad spectrum, single-channel PIT is observed. With all four nanobars present, the coupling between the dipole resonance of the nanobar 1, and the hybridization of the PIT resonance in nanobars 2-4, combine and interfere and for proper phase-coupled conditions the emergence of a double transparency window at 1079nm and at 1270nm is observed (black solid-line).

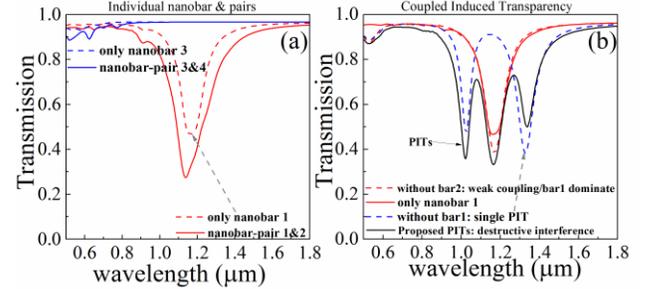

**Fig. 3.** Illustration of the resonant coupling leading to multispectral PIT. (a) Calculated transmittance spectrum for individual and nanobar pairs (see legend). (b) Calculated transmittance spectrum for the asymmetric topology shown in figure 1(b).

The calculated electric component $E_z$ amplitude and magnetic field $|H|$ distribution at resonances are shown in figure 4. As indicated from the results of figure 3, and 4, the transmission in the proposed system is manipulated by the interplay between the bright mode resonance (nanobar 1) and the coupled resonance of the closely arranged split-ring-resonator-like (SRR-like) nanostructures (nanobars 2, 3, and 4). For a wavelength of 1022nm the anti-bonding mode is primarily excited (fig. 4 (a, f)), while at 1165nm the strong dipolar resonance from nanobar1 dominates (fig. 4(c, h)). At the intermediate wavelength 1079nm both anti-bonding and bright modes coexist and interfere destructively to produce the first transparency window PIT A (fig. 4 (b,g)). Similarly, it is the interference between the hyper-bonding magnetic-resonance mode (fig. 4 (e, j)) and the bright mode that generates the second transparency window, PIT B (fig. 4 (d, i)).

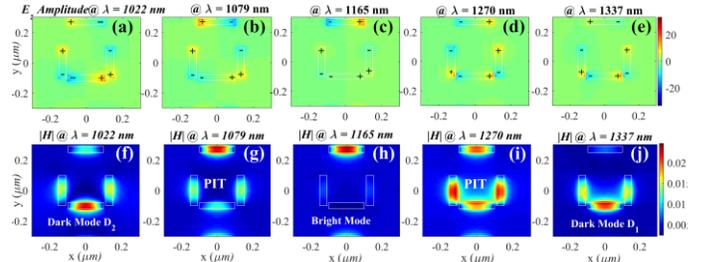

**Fig. 4.** $E_z$ field and magnetic field distribution at resonances calculated at the nanobar-substrate interface. (a, f) at 1022 nm (anti-bonding mode); (b, g) at 1079 nm (PIT A); (c, h) at 1165 nm ('bright mode'); (d, i) at 1270 nm (PIT B); (e, j) at 1337 nm (bonding mode).

The coupling between the bright and dark modes plays an important role for controlling the line shape of PIT transmission. The weak coupling and strong coupling boundaries are featured with a non-Hermitian degeneracy known as the exceptional point (EP) [23]. The transmission as a function of the coupled strength for varied values of $G$, and varied values $S_1$ are shown in figure 5. The results are illustrated in figure 5(b).

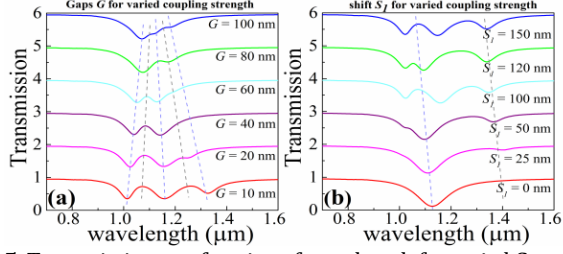

**Fig. 5.** Transmission as a function of wavelength for varied Gaps $G$ (a), and shift $S_1$ (b). The dashed lines are guide for the eye. The lines are plotted with an offset 1 for clarity.

The results in figure 5(a) indicate that, the gap between the nanobars relates to the strength of the overlap and the coupling between the fields of the different blocks in the SRR structure, where the dark modes excited and the PIT transmission are vastly modulated by $G$ (for larger $G$ values, the coupling degenerates and the PIT is gradually vanishing). On the other hand, the PIT effect vanishes and only a single dip can be seen (figure 5(b)) for a symmetric topology where $S_1 = 0$, (see figure 1(b)).

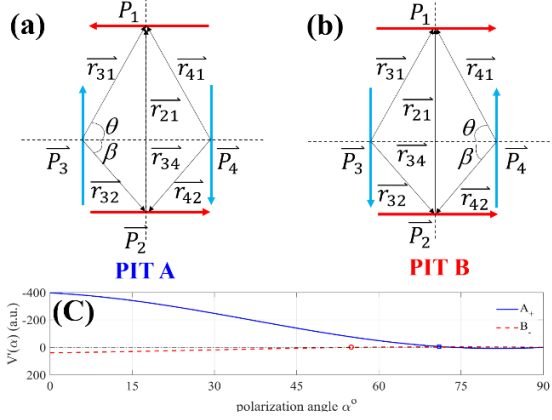

**Fig. 6.** Schematic of the interaction of in-plane electric dipoles moments for (a) PIT A, and (b) PIT B. (c) Calculated interactive energy $V'$ as a function of incidence polarization angle for the two degenerate resonances from *Eq (4)*.

Many previously reported multi-PIT systems achieve their properties through geometrical property manipulation and are consequently static. However, dynamic tuning of PITs is of great significance. In this work, we achieved versatile tailoring and tuning of the PIT channels by changing the incidence polarization orientation. To illustrate the physical mechanism, we calculate the interactive energy in the quasi-static approximation [24, 25]. The quasi-static interaction model is drawn (figure. 6) based on the distribution $E_z$ of peaks in PIT transmission. The quasi-static interactive electric energy between two dipoles $\vec{P_1}$ and $\vec{P_2}$ with their center-to-center vector $r$ is calculated as [24, 25].

$$V_e = \frac{1}{4\pi\varepsilon_0 r^3}(\vec{P_1} \cdot \vec{P_2} - 3(\vec{P_1} \cdot \hat{r})(\vec{P_2} \cdot \hat{r})) \quad (1)$$

The interaction of the dipole moments in the coupled nanoantenna system, where anti-parallel dipoles and electric dipole moments are considered for two coupled resonances are shown in Figure. 6. Considering the excitation conditions and properties of the bright and the dark modes, using the structural symmetry ($P_3 = P_4$), the electric interactive energy is calculated and simplified to be,

$$V_{eA,eB} = \frac{-3}{4\pi\varepsilon_0}\left\{\left(\frac{P_4 P_3}{3r_{43}^3} - \frac{P_3 P_1 r_{34}}{r_{31}^4}\sin\theta\right) \pm \left(\frac{P_2 P_1}{3r_{21}^3} - \frac{P_3 P_2 r_{34}}{r_{32}^4}\sin\beta\right)\right\}. \quad (2)$$

A measure of the interaction of incident light with spherical nanoparticle is given by the cross-section for dipole scattering [24, 26],

$$C_{sca} = \frac{k^4}{6\pi}\left|\frac{P}{E_0}\right|^2, \quad (3)$$

where $P$ is the dipole moment, $E_0$ is the applied field, and $k = 2\pi/\lambda$ denotes the wavevector. For nonspherical nanoparticles or nanorods, the scattering of electromagnetic waves with full mapping of the size and shape dependence of the dipolar plasmon resonance has been examined by several authors and should be treated accordingly [27-29]. Here, by employing the numerically calculated results in figure 2, and considering that the coupled excited dipole efficiency is polarization related, the corresponding modified electric interactive energy $V'$ can be written as,

$$V'_{eA,eB}(\alpha) = \frac{-3}{4\pi\varepsilon_0}\left[\begin{array}{c}(\frac{P_4 P_3}{3r_{43}^3}\sin^2\alpha - \frac{P_3 P_1 r_{34}}{2r_{31}^4}\sin\theta\sin 2\alpha) \pm \\ (\frac{P_2 P_1}{3r_{21}^3}\cos^2\alpha - \frac{P_3 P_2 r_{34}}{2r_{32}^4}\sin\beta\sin 2\alpha)\end{array}\right], \quad (4)$$

'+' for A, '−' for B.

The results in Fig. 6 (C) show the calculated electric interaction energy $V'$ for the two PIT resonance modes A and B. The degenerate modes for the proposed system and the shrinking of the interaction energy difference between the two are predicted for increased incidence polarization angles. Further increase of $\alpha$ leads to the domination of the total dipole resonance, and the disappearance of the PIT is intuitively predicted.

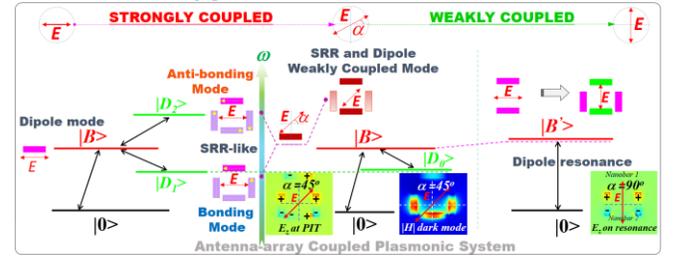

**Fig. 7.** Schematic of the system consisting of different meta-atom bright $|B\rangle$ and dark-mode $|D_{1,2}\rangle$ and the destructive interfering pathways for different polarization. The successively adjustable multiple PIT is presented.

Figure 7 shows the evolution of the coupling between bright-mode and dark-mode 'meta-atoms' for varied polarization. For horizontal polarization the bright (dipolar) and dark (SRR-like resonance) modes are strongly coupled giving rise to the antibonding and the bonding modes. The destructive interference of the pathways for the meta-atom energy level relaxation results in the double transparency-window. Similarly, for vertical polarization ($\alpha$=90°), since the dipole resonance scattering is about one order of magnitude stronger than the counterpart resonance of orthogonal polarization (figure 2), and the distance between nanobars 3 and 4 is large, weak coupling of the radiation from them is expected. Only bright resonance is expected and no PIT effect exists. For 45° polarization though, the concurrent excitation of the bright mode and the low-energy excited coupled SRR resonance, which are of the same order of magnitude, result in a single PIT window. The near-field distribution on resonance is demonstrated in Figure. 7.

To verify the theoretical predictions for the dynamic PIT, the transmission as a function of polarization is calculated in Figure 8 (a), where the evolution of the transparency-window numbers is shown. The bright mode with broader resonance linewidth centered at around 1100 nm destructively interferes with the excited dark modes featured with narrow resonance linewidths. The PIT effect can be switched between 'on' and 'off' state, and the activation of the different PIT channels (no-PIT, 1-Peak PIT, 2-Peak PIT) is realized by adjusting the polarization angle of the incident electric field as predicted by the quasi-static model and the theoretical analysis.

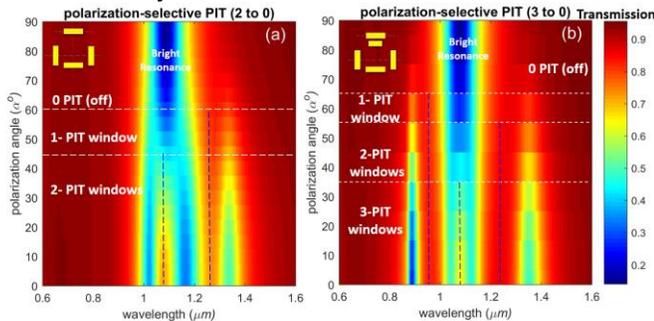

**Fig. 8.** Transmittance spectra as a function of incidence polarization angle ($S_1$ = 150 nm, with other parameters kept fixed as in figure 2). (a) Successively switchable shift of transmission windows from '2' to '0'. (b) Successively switchable shift of transmission windows from '3' to '0'. The dashed-lines are depicted as a guide for the eye. The horizontal dotted lines intuitively indicates the regions for transparency-windows. The unit structure is depicted in the upper-left.

It is noteworthy that the revealed mechanism implies that further transmission windows can be induced by introducing additional bright resonance unit. Further manifestation of triple-PIT is achieved with an extra-added dipole antenna adjacent to nanobar 1, the role of which is to enhance the coupling interaction with the bright mode. The results in figure 8 (b) demonstrate the effect of adding one more nanobar ($L$ = 140 nm, gap between nanobar 1 is fixed as 20 nm, and $S_1$ = 150 nm) to the system, where successively switchable shift of the numbers of the transparency-window is achieved (from '3' to '0'). Due to the introduction of the extra-nanobar, the resonance linewidth is slightly suppressed, and the resonance wavelength shows negligible shift from its original position.

In summary, we investigated the physical mechanism of plasmonic nano-antenna array structures that exhibit dynamic multispectral PIT resonance with successively polarization-switchable and tunable channels and controllable features. The general design approach and the multispectral construction mechanism with the polarization-excited nanoantenna element as the building blocks, provides way for efficient dynamic PIT design and manipulation. According to the scaling law, the induced transparency window can be readily extended to infrared frequencies or THz ranges [30]. Our findings exhibit the possibility of using simple nanostructure topology to generate dynamic configurable multispectral PIT, the results provide potential ways for promising applications such as plasmonic switching, dynamic metasurface, and multifunctional nanophotonics devices.